\documentclass{aa}
\usepackage{natbib,graphics}
\usepackage{times}
\bibpunct{(}{)}{;}{a}{}{,}

\begin{document}

\sloppy
  
\title{Spectral types of planetary host star candidates from OGLE III}
\author{S. Dreizler\inst{1}, S. Schuh\inst{1}, D. Homeier\inst{1}}

\offprints{Dreizler, dreizler@astro.physik.uni-goettingen.de}

\institute{Institut f\"ur Astrophysik, Georg-August-Universit\"at G\"ottingen,
           Friedrich-Hund-Platz~1, D--37077 G\"ottingen, Germany
}
\date{received; accepted}
\authorrunning{Dreizler et al.}
\titlerunning{Spectral types of OGLE III planetary candidates}
\abstract{The Optical Gravitational Lensing Experiment project has
  recently provided the OGLE III list of low-luminosity object
  transits from campaigns \#3 and \#4, reporting 40 new objects
  exhibiting the low-amplitude photometric eclipses expected for
  exoplanets. Compared to previous OGLE targets, these OGLE III
  candidates have been more restrictively selected and may contain
  low-mass planets.}{We have secured follow-up low-resolution
  spectroscopy for 28 candidates out of this list (and one from the
  OGLE Carina fields) to obtain an independent characterization of the
  primary stars by spectral classification and thus better constrain
  the parameters of their companions.}{We fed the constraints from
  these results back into an improved light curve solution. Together
  with the radius ratios from the transit measurements, we derived the
  radii of the low-luminosity companions. This allows us to examine
  the possible sub-stellar nature of these objects.}{Sixteen of the
  companions can be clearly identified as low-mass stars orbiting
  a main sequence primary, while 10 more objects are likely
  to have red giant primaries and therefore also host a stellar companion; 3
  possibly have a sub-stellar nature (R$\le 0.15\,$R$_\odot$).}{The
  planetary nature of these objects should therefore be confirmed by
  dynamical mass determinations.}

  \maketitle

\section{Introduction}
\label{sec:introduction}
The detection of planets outside our solar system has been a longstanding
goal of astronomy. After the first detections \citep{wolszczan:92,
mayor:95}, an intensive search with various methods began (see
\citealt{schneider:02} for an overview). Out of the more than 200
currently known planets, most 
have been detected with Doppler-velocity measurements of the planets'
(late type main sequence) host stars. With increasing precision, the
application of the Doppler-method could be extended to
Neptune-mass planets \citep[e.g.][]{2004A&A...426L..19S}.

Meanwhile, planet detections using alternatives to the Doppler method
have also been successfully applied. The first four microlensing
planets have been detected
\citep{2004ApJ...606L.155B,2005ApJ...628L.109U,2006Natur.439..437B,2006ApJ...644L..37G},
possible first direct images of extra-solar planets were published
\citep{2004A&A...425L..29C,2005A&A...438L..25C,2005A&A...438L..29C,2005A&A...435L..13N,2006ApJ...641L.141B},
and the number of detections due to transit searches increased to
fourteen
(
\citealt{2005ApJ...633..465S}; 
\citealt{2005A&A...444L..15B}; 
\citealt{2006ApJ...648.1228M}; 
\citealt{2006astro.ph..9369B}; 
\citealt{2006astro.ph..9335O}; 
\citealt{2006astro.ph..9688C} 
).

The transit method is of special interest, since it permits the derivation of
additional planetary parameters like the radius either indirectly via the
radius determination of the host star or directly via detection of 
the secondary eclipse as observed with the Spitzer Space Telescope
\citep{2005ApJ...626..523C,2005Natur.434..740D}. If combined with a
radial velocity variation measurement, the mass and mean density can be
determined, allowing us to obtain constraints for the planetary
structure. Furthermore, transiting systems allow us to investigate the
planets' atmospheres \citep{2002ApJ...568..377C,2004ApJ...604L..69V}. 

\begin{figure*}[ht]
\vspace{17.2cm}
\includegraphics{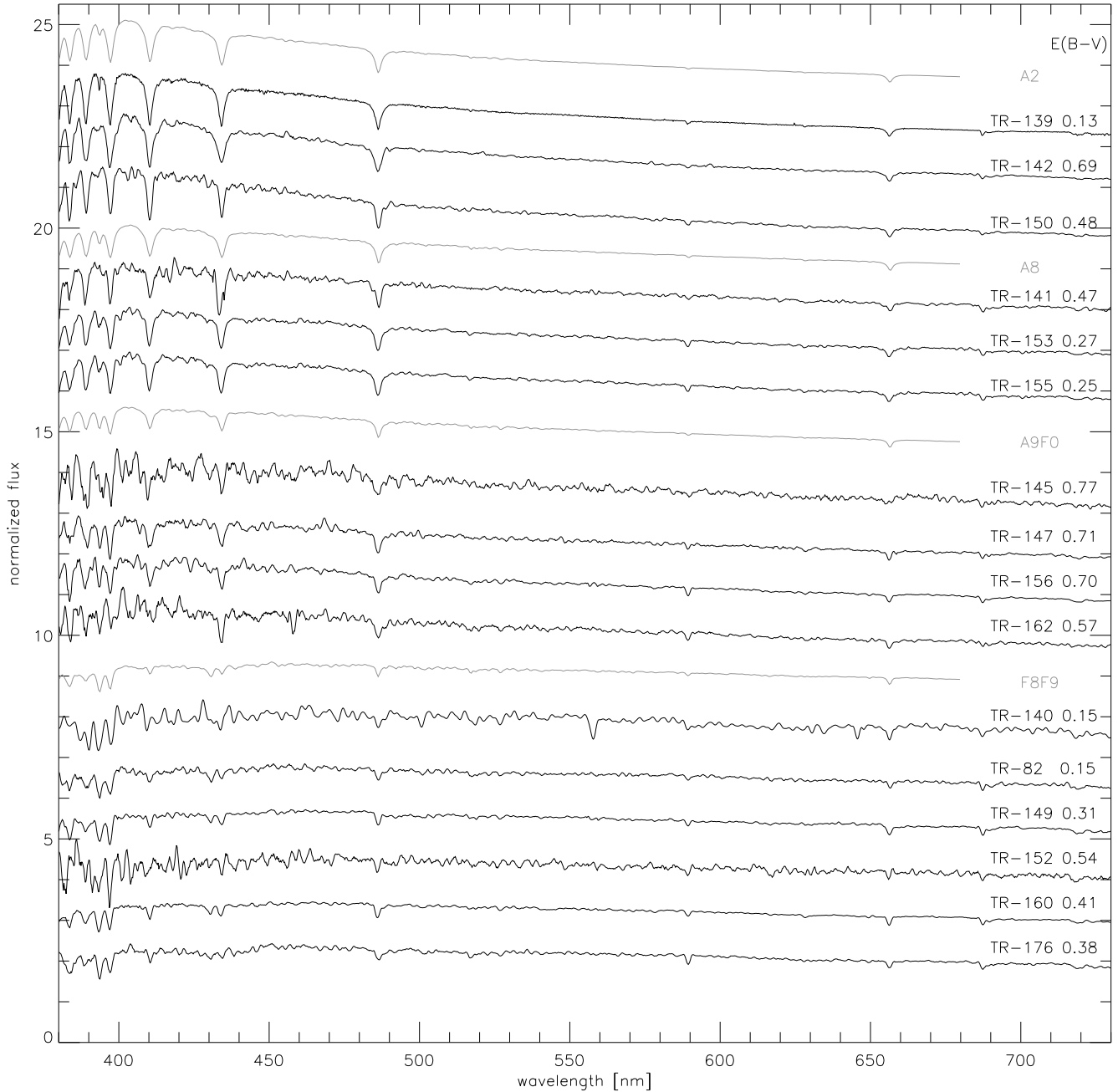}
\caption[]{Optical spectra of the early type target stars down to spectral
  type F9 together with template spectra (in grey) from
  \cite{silva:92} {\bf and the derived reddening}. See
  text for details.}  
\label{Fall}
\end{figure*}

\begin{figure*}[ht]
\vspace{15.2cm}
\includegraphics{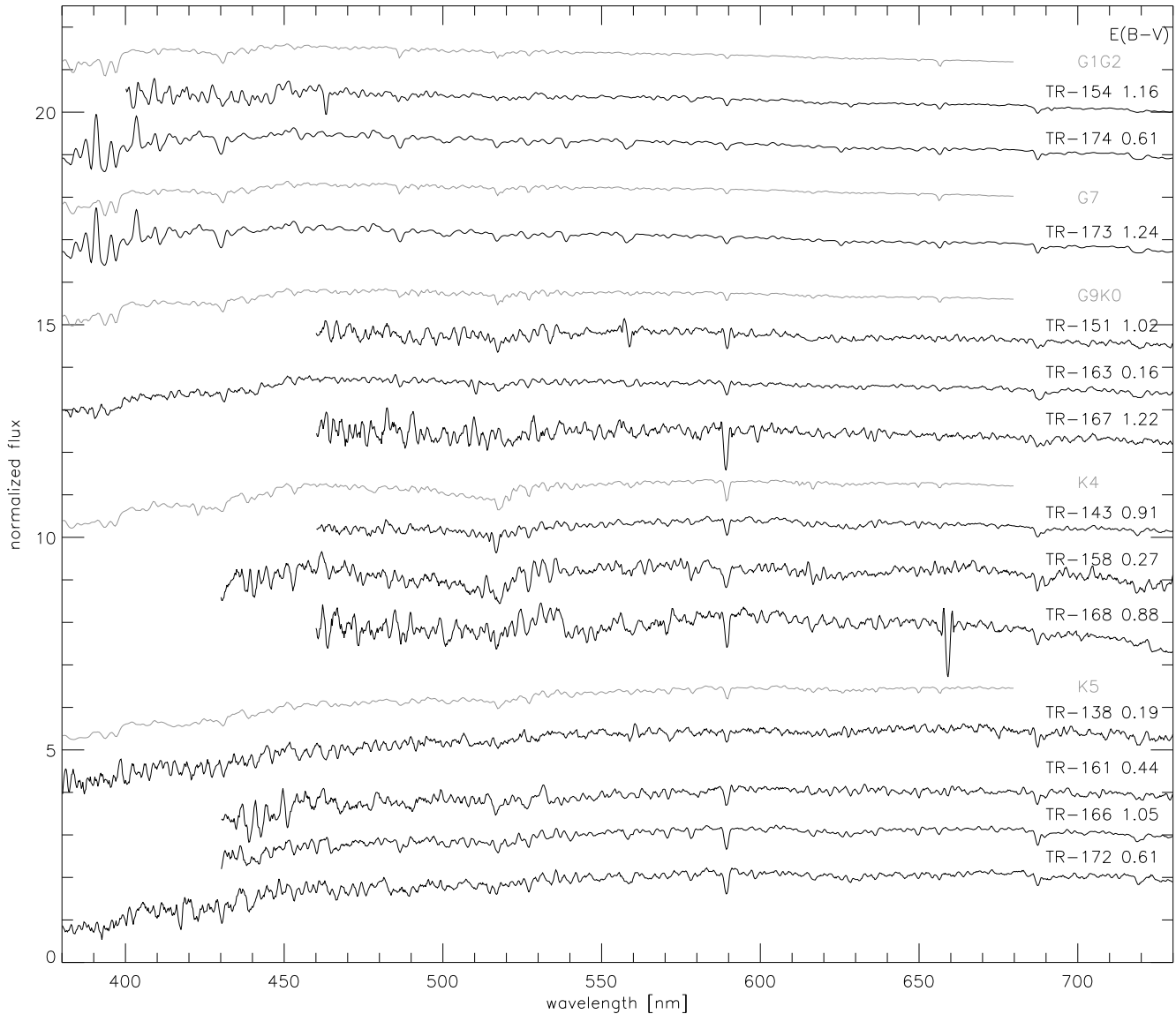}
\caption[]{Optical spectra of our late type target stars together with template
  spectra (in grey) from \cite{silva:92} {\bf and the derived reddening}. Depending on the brightness of 
  the object as well as on the observing conditions, spectra could only be
  extracted in a limited spectra range. See text for details.}  
\label{Fall2}
\end{figure*}

\begin{table*}[ht]
\caption{Results for our program stars: Derived spectral types and
  radii of the primary star (columns 2, 3). The transit depth
  from the OGLE light curve provides the companion radius R$_{\mathrm
  c}$ (columns 4 to 5) followed by
  the ellipsoidal variation $\delta$I$_{\mathrm e}$. The left hand value is
  derived from the OGLE light curve (``$<0.5$'' denotes cases where a
  formal fit results in amplitudes between 0.1 and 0.5 mmag), while the
  right hand value is from a \texttt{Nightfall} simulation of the
  corresponding system assuming a stellar companion (only values not
  in accordance with amplitudes in the observed light curves are
  given). \texttt{Nightfall} fits are given in columns 7 to 12
  followed by our final assessment. See text for details.} 
\begin{center}
\begin{tabular}{r|lc|ccr@{\,\,\,}l|cc|cc|cc|r@{}l}
\hline
$\#$& $\!$type$\!$ & $\!$R$_{\mathrm s}$/R$_{\odot}\!$ &
$\!$R$_{\mathrm c}$/R$_{\mathrm s}\!$ & $\!$R$_{\mathrm c}$/R$_{\odot}\!$ & 
\multicolumn{2}{c|}{$\delta$I$_{\mathrm e}$} & $\!$R$_{\mathrm s}$/R$_{\odot}\!$ &
$\!$R$_{\mathrm c}$/R$_{\odot}\!$ & i  & 
$\!$R$_{\mathrm c}$/R$_\odot\!$ & 3$^{\rm rd}$ light & $\!$R$_{\mathrm c}$/R$_{\odot}\!$  & \multicolumn{2}{c}{cat.} \\ 
 & & & & & \multicolumn{2}{c|}{[mmag]} & & & [deg] & & [\%] & &&\\
     & \multicolumn{2}{c|}{spectral type} & \multicolumn{4}{c|}{OGLE} &
\multicolumn{2}{c|}{\texttt{Nightfall} fit} &\multicolumn{2}{c|}{\texttt{Nightfall} fit}&\multicolumn{2}{c|}{\texttt{Nightfall} fit} &&\\ 
\hline
82 \rule{0cm}{3.7mm} & F8F9 & \hspace{-3mm}1.16$^{-0.1}_{+0.9}$ & 0.18 & 0.20$^{-.01}_{+.16}$ & 0.68 & 5.52 & 0.71$\pm .09$ & 0.10$\pm .03$ & $<80$     & 0.20$\pm .01$ & 70$\pm 10$   & \hspace{-3mm}0.29$\pm .02$ & S&  \\
138\rule{0cm}{3.7mm} & G9K0 & \hspace{-3mm}0.86$^{-0.1}_{+0.3}$ & 0.12 & 0.10$^{-.01}_{+.03}$ &      &      & 0.85$\pm .08$ & 0.10$\pm .05$ & 86$\pm 1$ & 0.10$\pm .01$ & 50$\pm 10$   & \hspace{-3mm}0.14$\pm .01$ &  &  \\
138\rule{0cm}{3.7mm} & K5   & \hspace{-3mm}0.72$^{-0.1}_{+0.1}$ & 0.12 & 0.09$^{-.01}_{+.01}$ & 0    &      & 0.67$\pm .04$ & 0.09$\pm .03$ & 89$\pm 1$ & 0.09$\pm .01$ & 25$\pm\,\,5$ & \hspace{-3mm}0.10$\pm .01$ & S&P \\
139\rule{0cm}{3.7mm} & A2   & \hspace{-3mm}2.12$^{-0.5}_{+1.7}$ & 0.17 & 0.35$^{-.08}_{+.28}$ & $<$.5& 1.28 & 1.70$\pm .22$ & 0.26$\pm .10$ & 80$\pm 1$ & 0.41$\pm .05$ & 80$\pm 10$   & \hspace{-3mm}0.73$\pm .20$ & S&  \\
140\rule{0cm}{3.7mm} & F8F9 & \hspace{-3mm}1.16$^{-0.1}_{+1.7}$ & 0.20 & 0.23$^{-.01}_{+.33}$ & 1.13 & 0.17 & 0.84$\pm .08$ & 0.17$\pm .05$ & 83$\pm 1$ & 0.42$\pm .02$ & 80$\pm 10$   & \hspace{-3mm}0.85$\pm .40$ & S&  \\
141\rule{0cm}{3.7mm} & A8   & \hspace{-3mm}1.58$^{-0.1}_{+1.7}$ & 0.13 & 0.21$^{-.01}_{+.22}$ & $<$.5&      & 1.47$\pm .50$ & 0.18$\pm .05$ & 83$\pm 4$ & 0.44$\pm .24$ & 15$\pm 15$   & \hspace{-3mm}0.20$\pm .01$ & S&  \\
142\rule{0cm}{3.7mm} & A2   & \hspace{-3mm}2.12$^{-0.5}_{+1.7}$ & 0.18 & 0.37$^{-.08}_{+.30}$ & 0    & 0.88 & 1.30$\pm .24$ & 0.18$\pm .05$ & 80$\pm 1$ & 0.51$\pm .06$ & 90$\pm\,\,5$ & \hspace{-3mm}0.87$\pm .13$ & S&  \\
143\rule{0cm}{3.7mm} & K4   & \hspace{-3mm}0.75$^{-0.1}_{+0.1}$ & 0.07 & 0.06$^{-.01}_{+.01}$ & $<$.5&      & 0.54$\pm .02$ & 0.06$\pm .04$ & 87$\pm 1$ & 0.08$\pm .01$ & 10$\pm 10$   & \hspace{-3mm}0.07$\pm .01$ &  &G \\
145\rule{0cm}{3.7mm} & A9F0 & \hspace{-3mm}1.52$^{-0.1}_{+1.7}$ & 0.13 & 0.20$^{-.01}_{+.22}$ & $<$.5&      & 1.79$\pm .12$ & 0.21$\pm .05$ & 89$\pm 1$ & 0.20$\pm .02$ &\,\,5$\pm\,\,5$&\hspace{-3mm}0.20$\pm .02$ & S&  \\
147\rule{0cm}{3.7mm} & A9F0 & \hspace{-3mm}1.52$^{-0.1}_{+1.7}$ & 0.17 & 0.25$^{-.01}_{+.27}$ & 0    &      & 1.57$\pm .35$ & 0.22$\pm .08$ & 82$\pm 5$ & 0.40$\pm .17$ & 70$\pm 20$   & \hspace{-3mm}0.57$\pm .23$ & S&  \\
149\rule{0cm}{3.7mm} & F8F9 & \hspace{-3mm}1.16$^{-0.1}_{+0.9}$ & 0.15 & 0.18$^{-.01}_{+.14}$ & 0.87 & 0    & 1.00$\pm .07$ & 0.16$\pm .02$ & 85$\pm 2$ & 0.24$\pm .06$ & 55$\pm 35$   & \hspace{-3mm}0.40$\pm .23$ & S&  \\
150\rule{0cm}{3.7mm} & A2   & \hspace{-3mm}2.12$^{-0.5}_{+1.7}$ & 0.07 & 0.14$^{-.03}_{+.11}$ & $<$.5&      & 1.98$\pm .07$ & 0.13$\pm .10$ & 83$\pm 2$ & 0.13$\pm .01$ & 70$\pm 20$   & \hspace{-3mm}0.28$\pm .11$ & S&  \\
151\rule{0cm}{3.7mm} & G9K0 & \hspace{-3mm}0.86$^{-0.1}_{+0.3}$ & 0.14 & 0.12$^{-.01}_{+.04}$ & 0    &      & 0.93$\pm .05$ & 0.11$\pm .02$ & 87$\pm 3$ & 0.13$\pm .02$ & 45$\pm 25$   & \hspace{-3mm}0.18$\pm .06$ &  &G \\
152\rule{0cm}{3.7mm} & F8F9 & \hspace{-3mm}1.16$^{-0.1}_{+0.9}$ & 0.13 & 0.15$^{-.01}_{+.12}$ & 0    &      & 1.52$\pm .13$ & 0.18$\pm .05$ & 86$\pm 3$ & 0.19$\pm .05$ & 60$\pm 10$   & \hspace{-3mm}0.28$\pm .08$ & S&  \\
153\rule{0cm}{3.7mm} & A8   & \hspace{-3mm}1.58$^{-0.1}_{+1.7}$ & 0.17 & 0.26$^{-.02}_{+.28}$ & 0.76 & 0.18 & 1.04$\pm .07$ & 0.17$\pm .04$ & 81$\pm 4$ & 0.54$\pm .27$ & 50$\pm 45$   & \hspace{-3mm}0.51$\pm .30$ & S&  \\
154\rule{0cm}{3.7mm} & G1G2 & \hspace{-3mm}1.05$^{-0.1}_{+0.6}$ & 0.11 & 0.11$^{-.01}_{+.06}$ & $<$.5&      & 1.23$\pm .08$ & 0.13$\pm .04$ & 84$\pm 3$ & 0.15$\pm .01$ & 35$\pm 35$   & \hspace{-3mm}0.20$\pm .08$ &  &G \\
155\rule{0cm}{3.7mm} & A6   & \hspace{-3mm}1.66$^{-0.1}_{+1.7}$ & 0.09 & 0.14$^{-.01}_{+.15}$ &      &      & 1.43$\pm .27$ & 0.13$\pm .06$ & 82$\pm 3$ & 0.28$\pm .10$ & 45$\pm 45$   & \hspace{-3mm}0.36$\pm .22$ &  &  \\
155\rule{0cm}{3.7mm} & A8   & \hspace{-3mm}1.58$^{-0.1}_{+1.7}$ & 0.09 & 0.14$^{-.01}_{+.15}$ & $<$.5&      & 1.53$\pm .11$ & 0.13$\pm .06$ & 84$\pm 5$ & 0.23$\pm .10$ & 50$\pm 40$   & \hspace{-3mm}0.35$\pm .22$ & S&  \\
156\rule{0cm}{3.7mm} & A9F0 & \hspace{-3mm}1.52$^{-0.1}_{+1.7}$ & 0.18 & 0.27$^{-.01}_{+.29}$ & 0    &      & 0.84$\pm .08$ & 0.15$\pm .02$ & 80$\pm 3$ & 0.40$\pm .05$ & 80$\pm 10$   & \hspace{-3mm}0.60$\pm .25$ & S&  \\
158\rule{0cm}{3.7mm} & K4   & \hspace{-3mm}0.75$^{-0.1}_{+0.1}$ & 0.13 & 0.11$^{-.01}_{+.01}$ & 0    &      & 0.39$\pm .02$ & 0.06$\pm .01$ & 87$\pm 1$ & 0.15$\pm .02$ & 15$\pm\,\,5$ & \hspace{-3mm}0.10$\pm .01$ & S&P:\\
160\rule{0cm}{3.7mm} & F6F7 & \hspace{-3mm}1.24$^{-0.1}_{+1.3}$ & 0.09 & 0.11$^{-.01}_{+.11}$ &      &      & 1.39$\pm .21$ & 0.12$\pm .05$ & 84$\pm 3$ & 0.17$\pm .05$ & 50$\pm 20$   & \hspace{-3mm}0.21$\pm .08$ &  &  \\
160\rule{0cm}{3.7mm} & F8F9 & \hspace{-3mm}1.16$^{-0.1}_{+0.9}$ & 0.09 & 0.10$^{-.01}_{+.08}$ & $<$.5&      & 1.12$\pm .09$ & 0.10$\pm .05$ & 85$\pm 2$ & 0.14$\pm .03$ & 25$\pm 20$   & \hspace{-3mm}0.15$\pm .04$ & S&P:\\
161\rule{0cm}{3.7mm} & G9K0 & \hspace{-3mm}0.86$^{-0.1}_{+0.3}$ & 0.07 & 0.06$^{-.01}_{+.02}$ &      &      & 1.54$\pm .08$ & 0.14$\pm .04$ & 90$\pm 0$ & 0.10$\pm .01$ & 10$\pm 10$   & \hspace{-3mm}0.11$\pm .01$ &  &  \\
161\rule{0cm}{3.7mm} & K5   & \hspace{-3mm}0.72$^{-0.1}_{+0.1}$ & 0.07 & 0.05$^{-.01}_{+.01}$ & 0    &      & 1.09$\pm .14$ & 0.10$\pm .05$ & 89$\pm 1$ & 0.06$\pm .01$ & 10$\pm 10$   & \hspace{-3mm}0.07$\pm .01$ &  &G \\
162\rule{0cm}{3.7mm} & A9F0 & \hspace{-3mm}1.52$^{-0.1}_{+1.7}$ & 0.10 & 0.15$^{-.01}_{+.17}$ & 0    &      & 1.12$\pm .17$ & 0.10$\pm .04$ & 81$\pm 4$ & 0.20$\pm .05$ & 50$\pm 40$   & \hspace{-3mm}0.31$\pm .17$ & S&  \\
163\rule{0cm}{3.7mm} & G7   & \hspace{-3mm}0.89$^{-0.1}_{+0.4}$ & 0.18 & 0.16$^{-.01}_{+.06}$ &      &      & 0.77$\pm .06$ & 0.12$\pm .01$ & 81$\pm 1$ & 0.15$\pm .01$ & 80$\pm 10$   & \hspace{-3mm}0.34$\pm .11$ &  &  \\
163\rule{0cm}{3.7mm} & G9K0 & \hspace{-3mm}0.86$^{-0.1}_{+0.3}$ & 0.18 & 0.15$^{-.01}_{+.05}$ &  0.5 & 2.22 & 0.73$\pm .04$ & 0.11$\pm .01$ & 81$\pm 1$ & 0.15$\pm .01$ & 70$\pm\,\,5$ & \hspace{-3mm}0.24$\pm .01$ & S&  \\
166\rule{0cm}{3.7mm} & G9K0 & \hspace{-3mm}0.86$^{-0.1}_{+0.3}$ & 0.10 & 0.09$^{-.01}_{+.03}$ &      &      & 1.09$\pm .06$ & 0.11$\pm .03$ & 90$\pm 0$ & 0.10$\pm .01$ & 25$\pm 25$   & \hspace{-3mm}0.13$\pm .02$ &  &  \\
166\rule{0cm}{3.7mm} & K5   & \hspace{-3mm}0.72$^{-0.1}_{+.01}$ & 0.10 & 0.07$^{-.01}_{+.01}$ &  0   &      & 0.97$\pm .03$ & 0.10$\pm .04$ & 90$\pm 0$ & 0.07$\pm .01$ & 25$\pm 25$   & \hspace{-3mm}0.09$\pm .01$ &  &G \\
167\rule{0cm}{3.7mm} & G9K0 & \hspace{-3mm}0.86$^{-0.1}_{+.03}$ & 0.14 & 0.12$^{-.01}_{+.04}$ &  0.83& 0    & 0.81$\pm .04$ & 0.11$\pm .03$ & 88$\pm 1$ & 0.14$\pm .02$ & 40$\pm 35$   & \hspace{-3mm}0.18$\pm .06$ &  &G \\
168\rule{0cm}{3.7mm} & K4   & \hspace{-3mm}0.75$^{-0.1}_{+0.1}$ & 0.11 & 0.08$^{-.01}_{+.04}$ &  0   &      & 0.80$\pm .03$ & 0.10$\pm .03$ & 90$\pm 0$ & 0.09$\pm .01$ & 40$\pm 30$   & \hspace{-3mm}0.13$\pm .03$ &  &G \\
172\rule{0cm}{3.7mm} & G9K0 & \hspace{-3mm}0.86$^{-0.1}_{+0.3}$ & 0.07 & 0.06$^{-.01}_{+.02}$ &      &      & 0.99$\pm .05$ & 0.08$\pm .04$ & 87$\pm 2$ & 0.09$\pm .01$ & 50$\pm\,\,5$ & \hspace{-3mm}0.12$\pm .02$ &  &  \\
172\rule{0cm}{3.7mm} & K5   & \hspace{-3mm}0.72$^{-0.1}_{+0.1}$ & 0.07 & 0.05$^{-.01}_{+.01}$ & 0.57 &  0.3 & 0.90$\pm .04$ & 0.07$\pm .03$ & 89$\pm 1$ & 0.06$\pm .01$ & 55$\pm 25$   & \hspace{-3mm}0.09$\pm .03$ &  &G \\
173\rule{0cm}{3.7mm} & G7   & \hspace{-3mm}0.89$^{-0.1}_{+0.4}$ & 0.13 & 0.12$^{-.01}_{+.05}$ & 0    &      & 0.69$\pm .07$ & 0.09$\pm .03$ & 83$\pm 2$ & 0.19$\pm .06$ & 35$\pm 35$   & \hspace{-3mm}0.13$\pm .07$ &  &G \\
174\rule{0cm}{3.7mm} & G1F2 & \hspace{-3mm}1.05$^{-0.1}_{+0.9}$ & 0.12 & 0.13$^{-.01}_{+.13}$ & $<$.5&      & 0.79$\pm .19$ & 0.08$\pm .05$ & 83$\pm 2$ & 0.19$\pm .04$ & 70$\pm 20$   & \hspace{-3mm}0.29$\pm .10$ &  &G \\
176\rule{0cm}{3.7mm} & F8F9 & \hspace{-3mm}1.16$^{-0.1}_{+0.9}$ & 0.14 & 0.16$^{-.01}_{+.16}$ & $<$.5&      & 1.26$\pm .14$ & 0.16$\pm .02$ & 84$\pm 3$ & 0.22$\pm .06$ & 35$\pm 35$   & \hspace{-3mm}0.28$\pm .10$ & S&  \\
\hline
\label{Tspec}
\end{tabular}
\end{center}
\end{table*}

This potentially large scientific output motivated the initiation of
several photometric monitoring projects to detect planetary
transits. Currently, OGLE (Optical Gravitational Lensing Experiment,
e.g.\
\citealt{2002AcA....52....1U,2002AcA....52..115U,2002AcA....52..317U,2003AcA....53..133U})
is the most successful one with five of the fourteen confirmed
transit planets out of 137 candidates. Additional 40 transiting
planetary/low-luminosity companions were announced by OGLE
\citep{2004AcA....54..313U}, increasing their total number of
candidates to 177.  These objects were extracted from a sample of
stars observed during two campaigns of several weeks of photometric
monitoring between February and July 2003 (and some additional
observations in 2004). In a sub-sample of 230\,000 stars with a
photometric accuracy better than 1.5\%, these 40 candidates exhibit
shallow ($\le$ 0.05 mag), flat-bottomed eclipses in their light curves,
indicating the presence of a transiting low-luminosity companion. The radii of the visible primaries
and of the invisible secondaries were derived from
the analyses of the light curves. Compared to previous
OGLE candidates, these are more restrictively selected. The candidate
list contains several (relatively) bright targets and several targets
with very shallow transits. Preliminary analyses of the light curves
by \citet{2004AcA....54..313U} indicate possible planets down to
Neptune size. The primary and companion radius estimates, derived from
the eclipse duration and the orbital period by adopting a main
sequence mass-radius relation, are based on the assumption of a
central passage. This can only be confirmed from light curve studies 
by detailed analysis of the transit shape
\citep{2003ApJ...585.1038S, 2005ApJ...627.1011T}, which is difficult
with the currently available photometric data.

Transit searches are a very promising method for detecting planets,
even from ground-based telescopes \citep{2005A&A...442..731G}. The
success of all ongoing and future transit searches depends on the
rejection of false positives caused by low-mass main sequence
companions, blends, and grazing eclipses. The necessity for such a
pre-selection can be seen from the large fraction of blends in the
sample of \citet[ 4 likely blends out of 5
candidates]{2003ApJ...597.1076K}. Based on OGLE candidate lists,
several selection techniques have been used to reduce the amount of
high-resolution spectroscopy for a final dynamical mass
determination. \citet{2005A&A...431..707G} use low resolution
spectroscopy in combination with IR photometry in order to exclude
giant primary stars and provide accurate effective temperatures and
radii of the OGLE-candidates. From low resolution spectroscopy,
\citet{2005AJ....130.1929H} have identified possible giant stars among the
candidates OGLE-TR-134 through OGLE-TR-137.

Up to now, no spectroscopic information of the primaries from the
latest OGLE-list \citep{2004AcA....54..313U} has been published. The
aim of this project therefore is to provide this information and to
decide about the nature of these low-luminosity companions.  The
spectral type provides a constraint for the primary radius which,
together with the transit light curve, permits a more accurate
determination of the radius of the unseen companion. Using data from
the SAAO 1.9\,m+SpCCD, we previously weeded out many A+M-star pairs
from the first OGLE list with this procedure
\citep{2002A&A...391L..17D}. The remaining objects with radii suitable
for sub-stellar objects can then be considered for further follow-up
to obtain a dynamical mass measurement for the unseen companion from
radial velocity variations.

We describe the observations, data reduction and the determination of
the spectral types of the primaries in Sect.\,\ref{sec:observations},
and present improved light curve solutions.  The results are discussed
in Sect.\,\ref{sec:discussion}.

\section{Observations, data reduction, and spectral types of the primary stars}
\label{sec:observations}

\subsection{Spectra}

We selected 28 candidates from the OGLE~III list of
\citet{2004AcA....54..313U}, with objects taken from both campaign \#3
(fields at RA=11\,h) and campaign \#4 (fields at RA=13\,h). All
objects with \mbox{m$_{I}<$16.35} were observed. Stars with entries in
the ESO Data Archive for high-resolution spectra were given a low
priority. Following the suggestion of \citet{2005A&A...438.1123P}, we
also added {\mbox OGLE-TR-82} to the target list (see
Table~\ref{Tspec}).  \par The spectra were obtained at the SAAO 1.9\,m
telescope between 2005/04/26 and 2005/05/02 using the spCCD Grating
Spectrograph equipped with a \mbox{266$\times$1798} SITe chip.  The
grating\,7 in combination with a slit width of 1.5" provided a
spectral resolution of 5\,\AA, and exposure times ranged from 1800\,s to
7200\,s depending on the observing conditions and the brightness of
the objects. Standard data reduction of these long-slit spectra was
performed using IDL and included bias subtraction, flat-field
correction, as well as wavelength calibration. Flux calibration was
done using spectra from the standard stars EG\,274 and LTT\,4364. Due
to the varying weather condition, the signal-to-noise ratio of the
spectra varies form 10 (OGLE-TR-168) to about 200
(OGLE-TR-139). Most spectra have a signal-to-noise ratio between 30
and 50.

The spectra were compared to the spectral library of \cite{silva:92}
without interpolation within the library. Both observed and template
spectra were normalized prior to comparison with a polynomial fit to
the continuum. The quality of the match between observed and template
spectrum was determined with an $\chi^2$ test (with a reduced $\chi^2$
ranging from 0.99 to 1.2 for the best fits) allowing us a classification
better than half a spectral class, sufficient for our further
analysis. For target stars with multiple spectra we obtained spectral
classifications within this error limit. Due to the absence of
sufficiently strong features for our spectral resolution and due to
the low signal-to-noise a discrimination between late G and mid K was
difficult so that we list the solutions for both possible
classifications.

In a first step, we restricted the the classification to luminosity
class V, because a discrimination between different luminosity classes
is too uncertain from our spectra. We also determined the interstellar
reddening assuming R$_{V}$=3.1 with the IDL routine {\em
ccm\_unred}. This spectral classifications of all objects and the
derived reddenings are displayed in Figs.\,\ref{Fall} and \ref{Fall2}
with the inverse polynomial function that we used to normalize the
template spectrum of the corresponding spectral type applied to the
observed spectrum. The presence of the companion could not be detected
from our data, neither from double lined spectra nor from the flux
distribution.

In a second step we checked our restriction to luminosity-class V with
a procedure adapted from \citet{2005A&A...431..707G}. The interstellar
extinction was modeled according to a simple homogeneous exponential
disc for distances from 0 to 1.5 kpc using $k_{0}=0.70$ and
$k_{0}=1.05$ . While \citet{2005A&A...431..707G} used IR photometry,
we took the I magnitudes from \citet{2004AcA....54..313U}, de-reddened
with the modeled extinction, and intrinsic colors corresponding to the
derived spectral type taken from \citet{2005ApJ...626..465R}. We could
then determine the limb-darkened angular diameter of the stars using
the results of \citet{2004A&A...426..297K}. The strong correlation of
mass and luminosity, as well as mass and radius on the main sequence,
allowed us to derive an effective temperature as a function of
distance (Eq.\,10 of \citealt{2005A&A...431..707G}). The effective
temperatures and the reddenings derived from the spectra should
therefore be consistent with this curve. This is not the case for
objects with extreme reddening even if we take uncertainties in the
brigtness, spectral class, and flux calibration with conservative
error bars into account. These stars are likely to be giants, marked
with {\em G} in the last column of Table\,\ref{Tspec}. For the further
evaluation, we however assume a main-sequence primary for all systems.

\begin{figure*}[htb]
\vspace{8.0cm}
\includegraphics{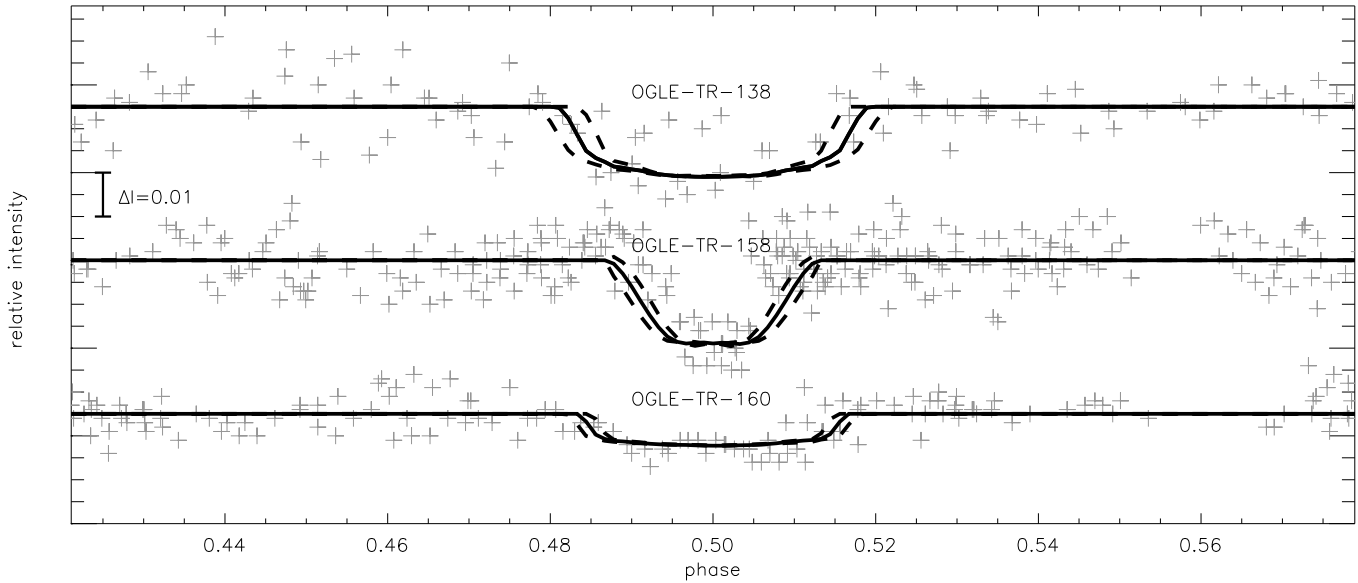}
\caption[]{OGLE light curves for the planetary companion candidates
  among our target stars with our new light curve solutions assuming a planet with one Jupiter
  mass and an inclination of 90\,$^\circ$. Solid line: Primary and
  companion radii from columns 3 and 5 of Table\,\ref{Tspec}, dashed
  lines: Lower and upper values for the radii as indicated in columns
  3 and 5.}
\label{Lall}
\end{figure*}

\subsection{Light curves}
We then used the derived spectral classes to estimate the stellar
radii of the primary stars
(Table~\ref{Tspec}, column 3) by interpolating the tabulated values from
\citet{allen}. The uncertainty range, also indicated in
Table~\ref{Tspec}, is calculated from zero-age and terminal-age
main-sequence models from \citet{1992A&AS...96..269S} and 
\citet{1999A&AS..135..405C} with solar metallicity. This reflects the
spread of interferometric radius determinations from
e.g. \cite{2004A&A...426..297K} for \mbox{$\alpha$ Cen A} and
\mbox{$\tau$ Ceti}, where the former is 20\% larger, the latter 10\%
smaller compared to the interpolated radius using the values from
\citet{allen} for a G2V or G8V star, respectively. The photometric
monitoring of \citet{2004AcA....54..313U} provides the brightness
variation during eclipses. Assuming a negligible radiation from the
secondary, a central passage in front of the primary, and no third
light contribution, this brightness variation is directly proportional
to the radius ratio squared. When multiplied by the primary radius it
yields the radius of the secondary (Table~\ref{Tspec}, column 5). Not
indicated in Table~\ref{Tspec} are the masses for the primary star,
which we derived from the tabulated values from \citet{allen},
\citet{1992A&AS...96..269S}, and \citet{1999A&AS..135..405C}
corresponding to the radii, and for the secondary object, which were
derived from evolutionary models for low-mass stars of
\citet{baraffe:98}, \citet{chabrier:00}, and \citet{baraffe:02}
assuming a low-mass star companion.

Using the derived radii and masses as initial start parameters, we fitted the
OGLE light curves with the close binary program
\texttt{Nightfall}\footnote{http://www.lsw.uni-heidelberg.de/$\sim$rwichman/Nightfall.html}
written by Rainer Wichmann. Best values and
uncertainties were derived from $\chi^2$-contours by varying the parameters. First, we fixed the
inclination to 90$^\circ$ and fitted the radii of both components
(Table~\ref{Tspec}, columns 7, 8). In a second step, we kept the
primary radius fixed at the spectroscopically-derived values and
fitted the orbital inclination, as well as the secondary radius
(columns 9, 10). Results generally agree with our
spectroscopically-derived values (Table~\ref{Tspec}). Finally, we also
investigated the possible presence of third light contribution
(columns 11, 12); however, the constraints from the light curves are
not very stringent in that case. The solution for the most promising
candidates, i.e. with radii equal or below 0.15\,R$_{\odot}$, are
shown together with the OGLE light curves in Fig.\,\ref{Lall} using
the values from columns 3 and 5 in Table~\ref{Tspec}.

We also investigated the ellipsoidal variation of the light curve out
of the eclipses (Table~\ref{Tspec}, column 6).  We estimated the
1$\sigma$ detection significance level to be about 0.3 mmag. {\mbox
OGLE-TR-140} shows the largest amplitude of 1.13 mmag. In this case,
the ellipsoidal variation can only be explained with a shorter orbital
period. Inspection of the original light curve indeed indicated a
period of 1.69665 days rather than 3.3933 days of
\citet{2004AcA....54..313U}. From our
\texttt{Nightfall}
simulation 
for the {\mbox OGLE-TR-140} system with a 1.69665 days orbital period
we derived an ellipsoidal variation with an amplitude of 1.16\,mmag, in
agreement with the observation, while the original period would result
in about a tenth of the observed amplitude. \object{OGLE-TR-149},
\object{OGLE-TR-153}, \object{OGLE-TR-167}, and \object{OGLE-TR-172}
also show lower amplitude in the simulations. As for {\mbox
OGLE-TR-140}, a shorter period might also be acceptable from the light
curve.  For \object{OGLE-TR-82}, \object{OGLE-TR-139},
\object{OGLE-TR-142}, and \object{OGLE-TR-163}, the amplitude from the
simulated light curves is clearly larger than from the observed
one. For \object{OGLE-TR-139} and \object{OGLE-TR-142}, twice the
orbital period would bring the simulations in agreement with
observations. For \object{OGLE-TR-82}, an at least a factor of three
longer period is required in order reduce the ellipsoidal variation to
the observed amplitude. The significance for the period of
\object{OGLE-TR-163} is very high due to 23 transit events. If we fix
the period and adjust the secondary mass in order to fit the
ellipsoidal variation, a brown-dwarf companion of 20-40 Jupiter masses
fits best.

\section{Discussion}
\label{sec:discussion}

Depending on the companion radii obtained from our investigations, we
defined three categories. (1) The 16 objects (including
\object{OGLE-TR-82}) in category {\em S} have companions of stellar
size, i.e. a radius larger than 0.15\,R$_\odot$ for all determined
values (columns 5, 8, 10, and 12 in Table\,\ref{Tspec}). (2) Regardless of
the analysis of the light curve, {\em G} classifies binaries that are
likely to have a giant primary star as derived from the spectral analysis
and therefore also have a main sequence companion. (3) {\em SP} are objects
with radii compatible with either small M stars or young sub-stellar
objects, i.e. a radius between 0.1 and 0.15\,R$_\odot$ (3
objects). Those with one of the values above the limit of
0.15\,R$_\odot$ are marked with a ``:''. There are no clear planetary
candidates, i.e. with radii below the stellar limit
(0.1\,R$_\odot$).

More follow-up observations enabling a dynamical mass determination
can finally confirm the nature of these objects. While objects from
category {\em S} and {\em G} can be excluded from future observations
aiming at planet detections, three deserve closer inspection:
\object{OGLE-TR-138}, \object{OGLE-TR-158}, \object{OGLE-TR-160}, and
possibly \object{OGLE-TR-163}, which either have very small stellar
companions like {\mbox OGLE-TR-122b} \citep{2005A&A...433L..21P} or
sub-stellar companions with large radii like HD\,209458b.

Several of the objects analyzed here are now subject to
high-resolution studies according to the ESO Data Archive. The near
future 
will therefore shed more light on the nature of several of our program
stars.

\begin{acknowledgements}
This paper uses observations made at the South African Astronomical
Observatory (SAAO, Run No.~163), kindly supported by D.\ Kilkenny. 
S.S.\ acknowledges a travel grant from the Deutsche
Forschungsgemeinschaft (DR-281/21-1). 
\linebreak
We acknowledge the use of the \texttt{Nightfall} program for the light
curve synthesis of eclipsing binaries written by Rainer Wichmann
(http://www.lsw.uni-heidelberg.de/$\sim$rwichman/Nightfall.html). 
\end{acknowledgements}

\end{document}